\documentclass[12pt,twoside]{article}
\usepackage{fleqn,espcrc1,amssymb}

\usepackage{graphicx}

\newcommand{\AmS}{{\protect\the\textfont2
  A\kern-.1667em\lower.5ex\hbox{M}\kern-.125emS}}
\newcommand{\be}{\begin{equation}}
\newcommand{\ee}{\end{equation}}
\newcommand{\bea}{\begin{eqnarray}}
\newcommand{\eea}{\end{eqnarray}}

\newcommand{\lton}{\mathrel{\lower.9ex
                  \hbox{$\stackrel{\displaystyle <}{\sim}$}}}

\title{
Discovery of Jet Quenching at RHIC and the Opacity of the Produced
Gluon Plasma} 

\author{P. L\'evai$^{\rm a}$, G. Papp$^{\rm b}$, G. Fai$^{\rm c}$, 
        M. Gyulassy$^{\rm d}$, 
        G.G. Barnaf\"oldi$^{\rm a}$, I. Vitev$^{\rm d}$ and 
        Y. Zhang$^{\rm c, \!\!\!}$
\address{
HAS KFKI Research Institute for Particle and Nuclear  Physics, 
PO Box 49, \mbox{Budapest 1525, Hungary} \\
$^{\rm b}$ HAS Research Group for Theoretical Physics, E\"otv\"os University,
P\'azm\'any P. 1/A, Budapest 1117, Hungary  \\
$^{\rm c}$ Physics Department, Kent State University, 
Kent OH 44242, USA \\
$^{\rm d}$ Physics Department, Columbia University, 
538 West 120th Street, New York, NY 10027, USA }}

\begin{document}

\maketitle
\begin{abstract}
The predicted quenching of jets in $A+A$ at RHIC energies
has been discovered by STAR and PHENIX in preliminary data
reported at this conference. We apply the GLV theory~\cite{glv2}
 of QCD radiative energy loss to estimate the
opacity, $L/\lambda_g$, of the gluon plasma produced in
$Au+Au$ collisions at 130 AGeV. We show that (in contrast
to the factor of two Cronin enhancement of $\pi^0$ found at the SPS
by WA98) the 
factor of 5-8 suppression of $p_T\sim 2-4$ GeV $\pi^0$ reported
by PHENIX~\cite{David00}
can be accounted for with an  effective static plasma opacity 
$L/\lambda_g\approx 3-4$. 
\end{abstract}

\section{Energy Loss of High Energy Jets}
The moderately high $p_{\rm T}< 6$ GeV 
hadron spectra in central Au+Au collisions,
especially $\pi^{\rm 0}$ production~\cite{David00},
reported at this conference,
reveal a 5-8 fold suppression compared to the 'expectation'
based on pQCD as we show below.
This suppression can be explained by the  energy loss
of moderate $p_T\sim 10$ GeV gluon mini-jets, the so-called
``jet-quenching'' effect predicted  a decade ago as
a signature of the production of high density matter with 
dense colored sources~\cite{gptw,mgxw92}. 
The observation  of jet quenching  is one of the major new
results from RHIC not observed before at lower SPS and AGS energies.
(See also talks of G. David, X.N. Wang, A. Dress, U.A. Wiedemann 
at this conference.)

Recently the QCD radiative energy loss was solved to all orders
in opacity $L/\lambda_g$ in~\cite{glv2,urs00}.    
Previous work  BDMS~\cite{bdms8}, based on asymptotic energy and thick target
approximations led to
$\Delta E_{\rm BDMS}={C_{\rm R}\alpha_{\rm s}}/{4} \cdot
{L^2\mu^2}/{\lambda_{\rm g}} \cdot \tilde{v}$,
where $C_{\rm R}$ is the color Casimir of the jet ($=N_{\rm c}$ for gluons),
and $\mu^2/\lambda_{\rm g}\propto \alpha_{\rm s}^2\rho$ is a  transport
coefficient of the medium proportional to the parton density, $\rho$.
The factor, $\tilde{v}\sim 1-3$  depends
logarithmically on $L$ and the mean free path, $\lambda_{\rm g}$,
namely $\propto \log (L/\lambda_{\rm g})$. Note that
it is the radiated gluon mean free path, $\lambda_{\rm g}$, that enters above.
\newpage

\begin{figure}[htb]
\vspace*{-1.2cm}
\begin{minipage}[t]{73mm}
\includegraphics[width=70mm]{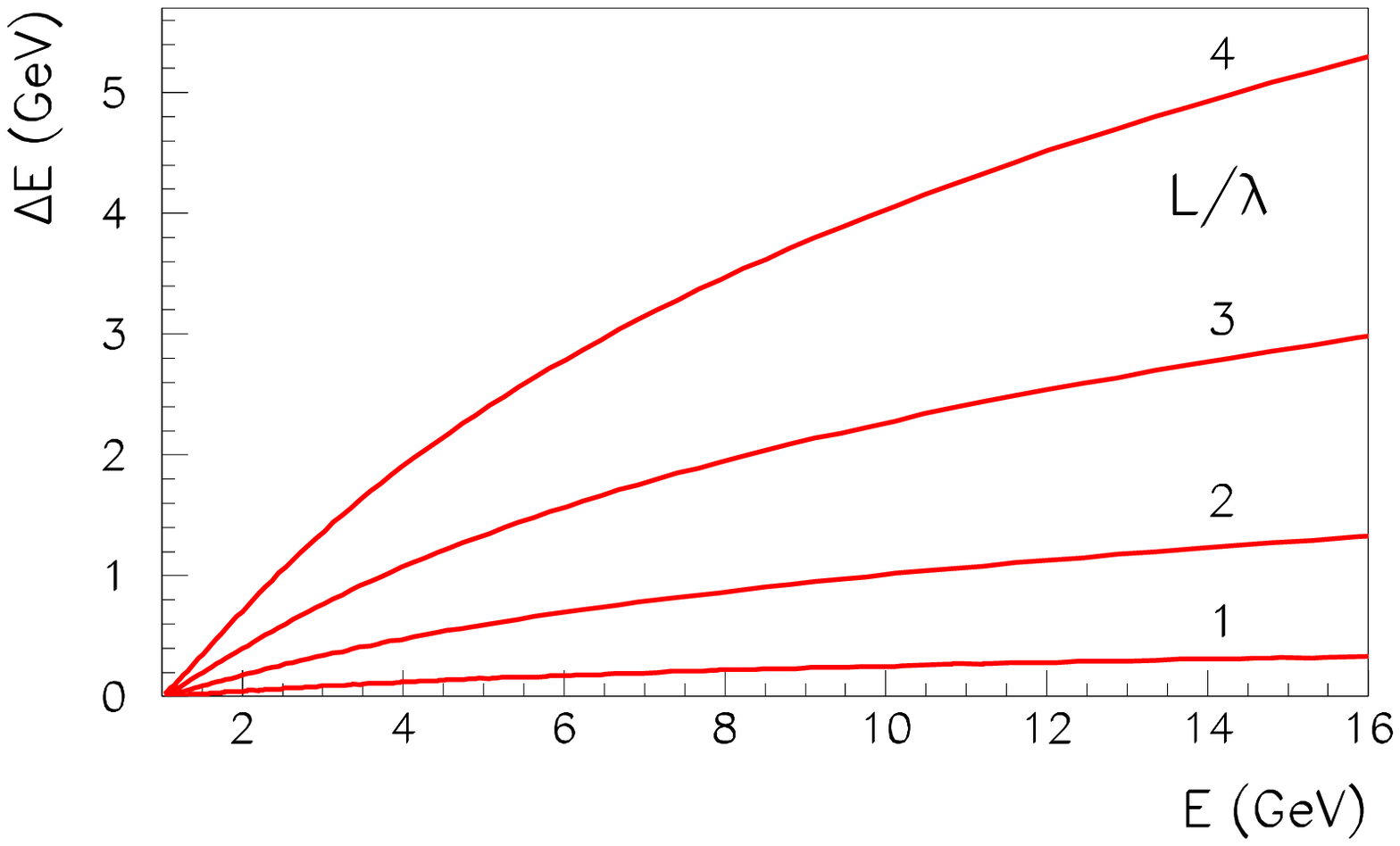}
\vspace{-40mm}
\caption{
Non-abelian energy loss of a gluon jet
calculated in the GLV picture~\cite{glv2}.
%At large energy the $\sim \log E$ behavior is reproduced.
}
\label{abse}
\end{minipage}
\hspace{\fill}
\begin{minipage}[t]{73mm}
\includegraphics[width=70mm]{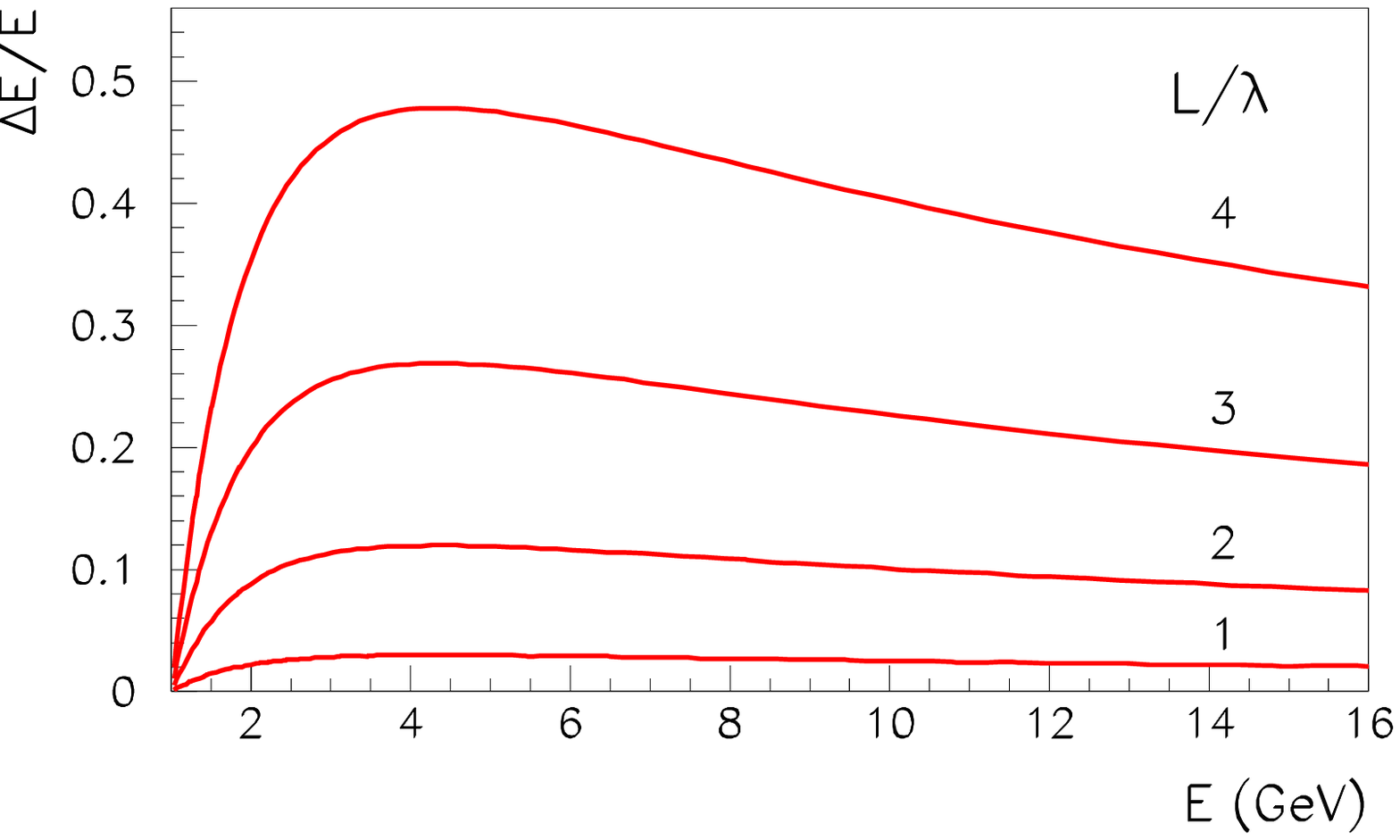}
\vspace{-40mm}
\caption{
The relative energy loss ($\Delta E/E$) is approximately constant
 at medium energy, $2 \leq E \leq 10$ GeV.
}
\label{rele}
\end{minipage}
\end{figure}
\vspace{-1.5cm}

In GLV~\cite{glv2} the theory was solved for the case of
``thin'' plasmas and  inclusion of finite kinematic effects.
Numerical results revealed that
the opacity series is strongly dominated by the first order term
\begin{eqnarray}
\Delta E_{\rm GLV}^{(1)}&=& \frac{2 C_{\rm R} \alpha_{\rm s}}{\pi}
\frac{E L}{\lambda_{\rm g}} \, \int_0^1 {\rm d}x
\int_0^{\rm k_{max}^2} \frac{{\rm d} {\bf k}^2_\perp}{{\bf k}^2_\perp}
  \int_0^{\rm q_{\max}^2}
\frac{ {\rm d}^2{\bf q}_{\perp} \, \mu_{\rm eff}^2 }{\pi
({\bf q}_{\perp}^2 + \mu^2)^2 } \cdot
\frac{ 2\,{\bf k}_\perp \cdot {\bf q}_{\perp}
  ({\bf k} - {\bf q})_\perp^2  L^2}
{16x^2E^2 \ + \ ({\bf k} - {\bf q})_\perp^4  L^2 }  \nonumber
\\[.5ex]
&= & 
\frac{C_{\rm R}\alpha_{\rm s}}{N(E)} \ 
\frac{L^2\mu^2}{\lambda_{\rm g}} \ \log \frac{E}{\mu} \,,
\label{dnx1}
\end{eqnarray}
where $N(E)=4$ in the asymptotic BDMS limit, but which is a rapidly 
increasing function of $E$ as the jet energy decreases in GLV.
A recent calculation~\cite{Zakh00} confirmed the validity of the GLV 
energy dependence.

We illustrate here the non-abelian energy loss
for an idealized  static plasma
with an average screening
scale $\mu=0.5$~GeV, $\alpha_{\rm s}=0.3$, and an average
gluon mean free path  $\lambda_{\rm g}=1$~fm.
The numerical results for the
first order energy loss $\Delta E$, taking into account the finite kinematic
bounds are displayed
in Figs. 1-2 for different opacities  ${\bar n} = L/\lambda_{\rm g}=1-4$.
As emphasized  in Ref.~\cite{glv2}
the first order  opacity result
 reproduces the characteristic BDMS quadratic dependence of the energy loss
on $L$.

However, the most important feature of Fig. 2 relative to
observable consequences of energy loss is that
unlike in the asymptotic BDMS case, where $\Delta E/E\propto
1/E$ increases rapidly with decreasing energy, the structure of the first order
expression as well as finite kinematic
bounds in the GLV expression lead to an approximate {\em linear}
energy dependence of $\Delta E$ in
the energy range, $E=2-10$ GeV and a slow
$\propto \log (E)/E$ decease at much higher energies.

\begin{figure}[htb]
\vspace*{-0.8cm}
\begin{minipage}[t]{75mm}
\includegraphics[width=70mm]{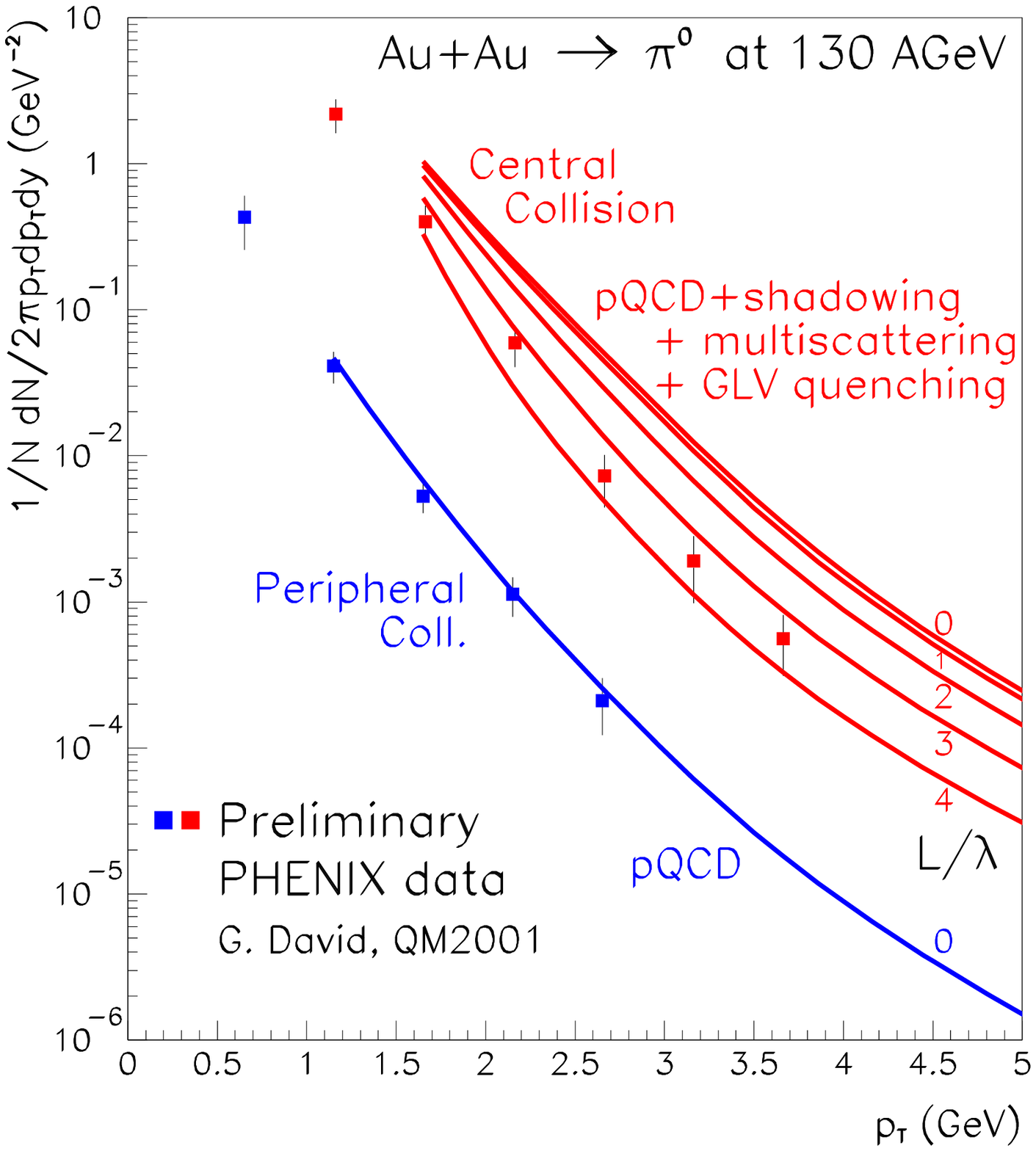}
\vspace{-12mm}
\caption{
Pion production in Au+Au collision
including jet quenching with opacity $L/\lambda=1,2,3,4$.
Data are from Ref.~\cite{David00}.
}
\label{abspi0}
\end{minipage}
\hspace{\fill}
\begin{minipage}[t]{75mm}
\includegraphics[width=70mm]{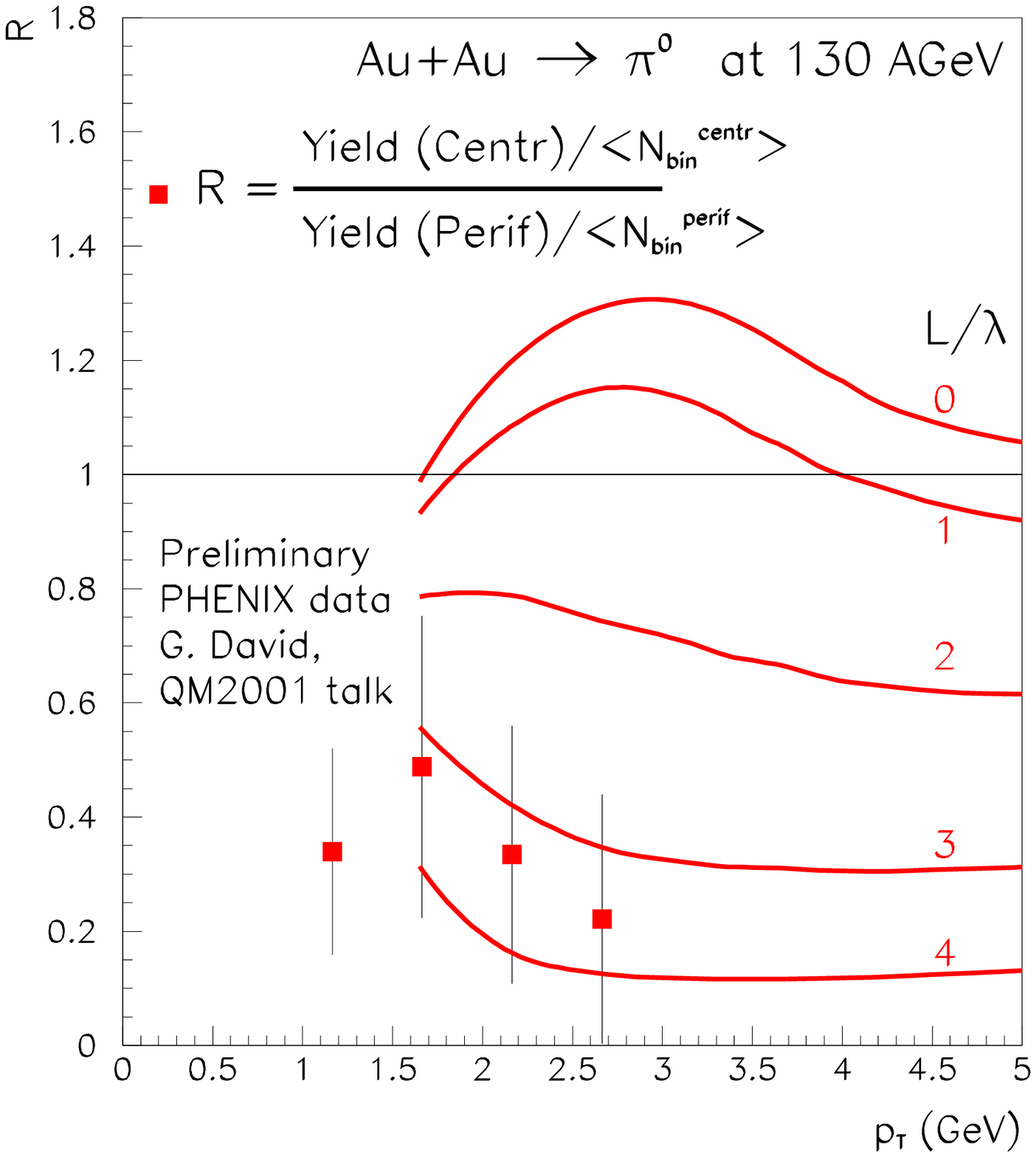}
\vspace{-12mm}
\caption{The ratio of the central to
the periferial pion yields (normalized by the number of binary collisions,
857 and 5.5).
}
\label{relpi0}
\end{minipage}
\end{figure}

%\vspace{-12mm}
\section{Pion Spectra in Au+Au Collisions from Perturbative QCD}

In order to investigate
the influence of the GLV  energy-dependent radiative energy loss
on  hadron production, we apply a pQCD based description of
Au+Au collisions, including energy loss prior
to hadronization.
In Refs.~\cite{PLF00,kapiquench} we displayed
how pQCD calculations can reproduce
the experimental data in p+p $\longrightarrow \pi$+X
reactions. Our calculations
incorporate the parton transverse momentum
(``intrinsic $k_{\perp}$'') via a Gaussian transverse momentum
distribution (characterized by
the width $\langle k_{\perp}^2 \rangle $)~\cite{PLF00,XNWint,QM01pg}. 

In the pQCD calculation of nuclear collisions shadowing 
will modify the parton distribution functions (pdf-s) in the
nucleons inside the nucleus. These modified pdf-s are denoted by
$f_{\rm a/A}(x_{\rm a},k_{\perp a},Q^2)$ 
and $f_{\rm b/A}(x_{\rm b},k_{\perp b},Q^2)$.
The finite size of the colliding nuclei implies an impact parameter 
dependence. The influence of
hadronic multiscattering is investigated in detail in Refs.~\cite{PLF00,QM01pg}.

Considering the energy loss effect, 
jet quenching reduces  the energy
of the jet before fragmentation. We concentrate on 
mid-rapidity ($y_{\rm cm}=0$),
where the jet transverse
momentum before fragmentation is shifted by the energy loss~\cite{WaHu97},
$p_{\rm c}^*(L/\lambda) = p_{\rm c} - \Delta E(E,L)$. This
shifts the $z_{\rm c}$ parameter in the fragmentation function
of the integrand (\ref{fullaa})
to $z_{\rm c}^* = z_{\rm c} /(1-\Delta E/p_{\rm c})$.

With these elements the invariant cross section of hadron
production in central $A+A$ collision is given by
\begin{eqnarray}
\label{fullaa}
&&E_{\rm h}\frac{{\rm d}\sigma_{\rm h}^{\rm AA}}{{\rm d}^3p} 
        =\int {\rm d}^2 b \  {\rm d}^2 r\ t_{\rm A}({\vec b}) 
         t_{\rm B}({\vec b} - {\vec r})
        \sum_{\rm abcd}\!
        \int\!\!{\rm d}x_{\rm a} {\rm d}x_{\rm b} {\rm d}z_{\rm c}
        {\rm d}^2k_{\perp,{\rm a}} {\rm d}^2k_{\perp,{\rm b}} \cdot
        \nonumber \\
        && \ \ \ \  f_{\rm a/A}(x_{\rm a},k_{\perp,{\rm a}}({\vec b}),Q^2) 
         f_{\rm b/A}(x_{\rm b},k_{\perp,{\rm b}}({\vec b} - {\vec r}),Q^2)\
         \frac{{\rm d}\sigma}{{\rm d}{\hat t}} \frac{z^*_{\rm c}}{z_{\rm c}}
   \frac{D_{\rm h/c}(z^*_{\rm c},{\widehat Q}^2)}{\pi z_{\rm c}^2} \,
    {\hat s} \delta({\hat s} + {\hat t} + {\hat u}) \,\, ,
\end{eqnarray}
where upper limit of the impact parameter integral is
 $b_{\rm max}= 4.7$ fm for 10 \% central Au+Au collisions.
Here $t_{\rm A}(b)$ is the usual (Glauber) thickness function.
The factor $z^*_{\rm c}/z_{\rm c}$ appears because of  the in-medium
modification of the fragmentation function~\cite{WaHu97}.
Thus, the invariant cross section
(\ref{fullaa}) will depend on the average
opacity or collision number, ${\bar n} = L/\lambda_{\rm g}$.
The calculated spectra for pions  are displayed
for ${\bar n}=0,1,2,3,4$ in Fig. 3. Fig. 4. shows
their ratios to the non-quenched spectra at ${\bar n}=0$.
We note that in contrast to previous energy independent
estimates for the energy loss, the GLV energy-dependent
energy loss leads to constant suppression of the high $p_T$
domain in agreement with the preliminary data.
The  peripheral collisions are consistent with a rather small opacity
in contrast to central collisions as expected.
 The ratio of central to peripheral PHENIX~\cite{David00} data
 shown in Fig. 4
clearly reveals that jet quenching at RHIC overcomes the Cronin
enhancement at zero (final state) opacity. This is in stark contrast to
data at SPS energies, where WA98 found no evidence for quenching
in $Pb+Pb$ at 160 AGeV but a factor of two Cronin enhancement 
(see talk of X.N. Wang).

In summary, 
Figs. 3-4 indicate that an
effective static plasma opacity $L/\lambda = 3-4$ 
is sufficient to  reproduce the striking jet quenching pattern observed
at RHIC. In the talk of Wang, a rather small constant $dE/dx\approx
0.25$ GeV/fm was also found to be consistent with the data.
However, it is important to emphasize that these effective {\em static}
plasma opacities and parameters hide the underlying rapid
dilution of the plasma due to expansion.
The GLV formalism including the kinematic constraints
at first order has been further generalized to include effects of 
expansion in~\cite{gvw}. 
It was found in~\cite{gvw} that  the inclusion of longitudinal
expansion modifies the static plasma results in such a way that the 
moderate static plasma opacity actually
 implies that the produced  mini-jet 
plasma rapidity density may have reached $dN_g/dy\sim 500$.
If confirmed by further measurements and theoretical
refinements, jet quenching may have already provided
the first evidence  
that initial parton densities 
on the order of 100 times nuclear matter density are produced at RHIC.
The full analysis of flavor composition, shape, and azimuthal moments
of the high $p_T$ spectra appears to be 
a promising diagnostic probe of the
evolution of the gluon plasma produced at RHIC.

{\bf Acknowledgments:}
We thank G. David for providing the PHENIX data and for discussions.
This research has been supported by the
Bolyai Fellowship of the Hungarian Academy of Sciences 
(P. L.), by the Hungarian OTKA grants  T032796, T034842
by the US-Hungarian Joint Fund MAKA 649/1998,
by BCPL Bergen, Norway, 
by the U.S. DOE under DE-FG02-86ER40251, DO-AC03-76SF00098, 
DE-FG02-93ER40764 and by the U.S. NSF under 
INT-0000211 (MTA-038).

\end{document}